\documentclass[aps,article,floats,twocolumn,superscriptaddress,english]{revtex4-1}
\usepackage{amssymb}
\usepackage{graphicx}
\usepackage{amsmath}
\usepackage{amsfonts}
\usepackage[T1]{fontenc}
\usepackage[latin9]{inputenc}
\usepackage{array}
\usepackage{multirow}
\usepackage{color}
\usepackage{esint}
\usepackage{bm}
\usepackage{color}
\usepackage{bbm}
\usepackage{hyperref}
\usepackage{titlesec}
\usepackage{soul}
\usepackage{mathtools}
\begin{document}

\title{Creation of a tweezer array for cold atoms utilizing a generative neural network}

\author{Zejian Ren}
\thanks{These authors contributed equally.}
\affiliation{Microelectronics Thrust, The Hong Kong University of Science and Technology (Guangzhou), Guangzhou, China}

\author{Xu Yan}
\thanks{These authors contributed equally.}
\affiliation{Department of Physics, The Hong Kong University of Science and Technology, Clear Water Bay, Kowloon, Hong Kong, China}

\author{Kai Wen}
\affiliation{Microelectronics Thrust, The Hong Kong University of Science and Technology (Guangzhou), Guangzhou, China} 

\author{Huijin Chen}
\affiliation{Microelectronics Thrust, The Hong Kong University of Science and Technology (Guangzhou), Guangzhou, China} 

\author{Elnur Hajiyev}
\affiliation{Department of Physics, The Hong Kong University of Science and Technology, Clear Water Bay, Kowloon, Hong Kong, China}

\author{Chengdong He}
\affiliation{Department of Physics, The Hong Kong University of Science and Technology, Clear Water Bay, Kowloon, Hong Kong, China}

\author{Gyu-Boong Jo}
\email[Corresponding author email: ]{gbjo@ust.hk}
\affiliation{Microelectronics Thrust, The Hong Kong University of Science and Technology (Guangzhou), Guangzhou, China} 
\affiliation{Department of Physics, The Hong Kong University of Science and Technology, Clear Water Bay, Kowloon, Hong Kong, China}

\begin{abstract}
Optical tweezers have become essential tools for dynamically manipulating objects, ranging from microspheres or biological molecules to neutral atoms.  In this study, we demonstrate the creation of tweezer arrays using a generative neural network, which allows for the trapping of neurtal atoms with tunable atom arrays. We have successfully loaded cold strontium atoms into various optical tweezer patterns generated by a spatial light modulator (SLM) integrated with generative models. Our approach shortens the process time to control the SLM wtih minimal time delay, eliminating the need for repeated re-optimization of the hologram for the SLM.
\end{abstract}
\maketitle
\newpage

\vspace{10pt}

\paragraph*{\bf Introduction} An optical tweezer, a highly focused laser beam,  have found applications in a wide range of fields, including biology, chemistry, and material science~\cite{minowa2022optical}. In recent years, optical tweezers have also been widely exploited in quantum science~\cite{Kaufman.2021}, where they have enabled the control and manipulation of individual atoms~\cite{endres2016atom,Kim.2016} and molecules~\cite{Cairncross.2021} in the quantum regime with unprecedented precision. The ability to control quantum systems at the individual level is crucial for the development of quantum technologies, such as quantum computing~\cite{ebadi2022quantum}, quantum simulation~\cite{Browaeys.2020} and quantum sensing~\cite{degen2017quantum}. 

To date, the creation of tweezer arrays using the spatial light modulator (SLM) has been mostly achieved through classical algorithms like the Gerchberg-Saxton (GS) algorithm~\cite{gerchberg1972practical} and its variations~\cite{kim2019large}. However, these methods often face a trade-off between the quality of reconstruction and the time required for calculations. Moreover, the time-consuming nature of these algorithms limits their potential application in quantum control with dynamic shot-to-shot re-configuration, as the quantum system's coherence time is finite~\cite{zhang2016trapped}. Recently, the emergence of artificial intelligence presents a new opportunity to optimize control~\cite{Wigley.2016,Milson.2023}, analysis~\cite{Zhang.20191q,Rem.2019,Bohrdt.2019,ness2020single,Zhao.2021,Zhao.2022} and monitoring~\cite{chen2023magnetic}  of quantum experiments, while also providing a new approach to holography~\cite{he2016deep}. In this work, we utilize the power of artificial intelligence and propose a simple deep learning approach that utilizes generative neural networks to generate atom arrays on demand based on the U-net model~\cite{ronneberger2015u}. Our work offers an effective way to advance quantum control through the use of machine learning, which holds the potential to drive dynamic control in optical-tweezer-based systems.

\begin{figure}
\includegraphics[width=1.\linewidth]{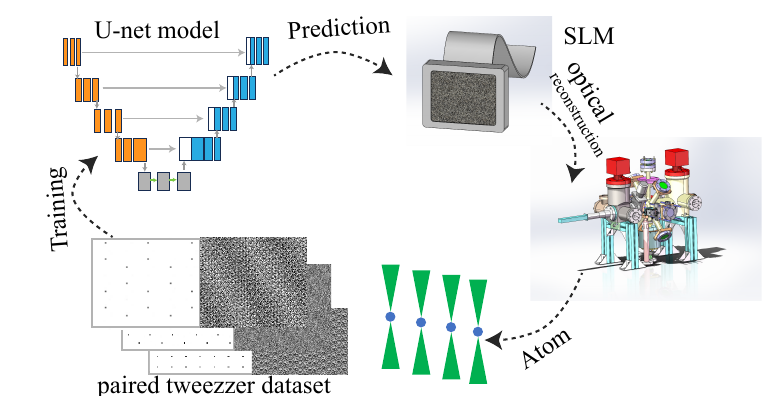}
\caption{ \textbf{Overview of the scheme for producing an atom array based on deep learning} To trap $^{88}$Sr atoms, a tweezer array is utilized. The Spatial Light Modulator (SLM) is controlled by a generative neural network that is based on the U-net model. This generative model is trained using the tweezer dataset. The high level of control provided by this setup allows us to create precise atomic configurations.}
	\label{fig1}
\end{figure}

In the typical setup, a static optical pattern is generated by utilizing the SLM, resulting in the formation of an optical tweezer array for neutral atoms. To achieve this, a phase-only hologram (POH) needs to be loaded into the SLM. This approach allows for the generation of arbitrary tweezer geometries, providing flexibility and control. However, there is a trade-off between the calculation time and the reconstruction time associated with this method.

One of the challenges with this technique is the need to re-optimize the phase-only hologram whenever a new tweezer pattern is required. This re-optimization step can be time-consuming and may hinder the real-time operation of the atom array system. To address this limitation, our objective is to leverage a trained generative neural network. By employing this neural network, we aim to bypass the re-optimization process and realize the generation of real-time atom arrays. By utilizing the trained generative neural network, we anticipate significant improvements in the efficiency and speed of generating atom arrays. This approach would eliminate the need for repeated re-optimization of the phase-only hologram, saving valuable time and resources. Ultimately, our goal is to enable the realization of real-time atom arrays with enhanced flexibility and control, opening up new possibilities for research and practical applications in the field of neutral atom manipulation.

\vspace{15pt}
\paragraph*{\bf General idea} 
Fig.~\ref{fig1} illustrates the basic concept of using a neural network model to create a tweezer atom array. Initially, we create two sets of data comprising the POH and the corresponding light pattern produced by the SLM. Random tweezer patterns with uniform light intensity are generated to form the datasets. We obtain the light pattern from a monitoring CCD camera while each corresponding POH is created by the GS algorithm~\cite{gerchberg1972practical}. Then, the designated dataset is sequentially fed as input into the neural network based on the U-net model~\cite{ronneberger2015u}. We train the generative neural network with the datasets consisting of paired matrix and phase data. Finally, the trained network is subsequently utilized to predict the POH for generating a tweezer pattern. The resultant POHs predicted by the generative model are then transferred onto SLMs for trapping neutral atoms in the experiment. In the experiment, we test the method by loading cold strontium atoms from a magneto-optical trap to tweezer traps achieved by the SLM with generative models~\cite{wen.2024}. We successfully generated different configurations of tweezers, with an emphasis on creating specific patterns without experiencing any significant time delay.

\begin{figure}
	\includegraphics[width=0.99\linewidth]{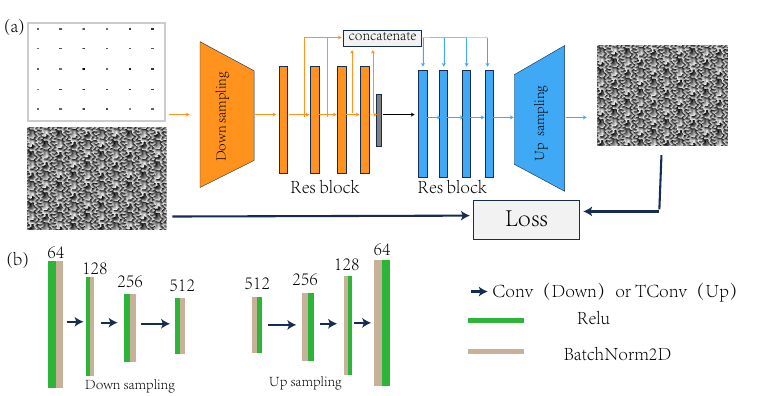}
	\caption{ {\bf Diagram of the neural network based on the U-net model} (a) The architecture of generative neural network accepts a pair of images, followed by the U-net model.  (b) Down sampling and Up sampling block architecture. The convolution (Conv) layer is used in down sampling block to increase the channel number while the transpose convolution (TConv) layer is added to the Upsampling block in order to reconstruct the phase information. }\label{fig2}
\end{figure}

\vspace{15pt}

\paragraph*{\bf Hologram dataset of the atomic array}
As the first step to facilitate the training of the generative model, we introduce a large-scale hologram dataset consisting of 2000 pairs of tweezer  images and corresponding holograms for the SLM (Holoeye; PLUTO-2.1-VIS-130). Since the tweezer is usually formed in a far-field regime, we use the traditional Fraunhofer zone method to create the dataset for training our model, resulting in the optical reconstruction on the object plane. The diffraction process in the GS algothrim is calculated according to the Fraunhofer diffraction~\cite{Hecht.1989}.
$$\begin{aligned}
	\hat{I}\left(x,y\right)& =|\hat{\mathbf{C}}(\boldsymbol{x},\boldsymbol{y})|^{2}=|\mathcal{F}\left\{\exp[\mathrm{i}\boldsymbol{\Phi}\left(\boldsymbol{x}_{0},\boldsymbol{y}_{0}\right)]\right\}|^{2}  \\
	&=|{\mathcal F}\left\{\exp[\mathrm{i}\boldsymbol{\varphi}\left(\boldsymbol{x}_{0},\boldsymbol{y}_{0}\right)]\cdot\exp\left[\mathrm{i}\frac{\pi}{\lambda d}\left(\boldsymbol{x}_{0}^{2}+\boldsymbol{y}_{0}^{2}\right)\right]\right\}|^{2},
\end{aligned}$$  
where $x,y,{x}_{0},{y}_{0}$ represent the coordinates on the object plane and the hologram plane, $\mathbf{C}(\boldsymbol{x},\boldsymbol{y})$ represents the complex amplitude distribution on the object plane, $\mathcal{F}$ denotes the Fourier transform, $\boldsymbol{\varphi}$ is the phase generated by the network. In this regime, the condition  $d\lambda\gg\boldsymbol{x}_{0}^{2}+\boldsymbol{y}_{0}^{2}$ is well-satisfied resulting in $exp\left[\mathrm{i}\frac{\pi}{\lambda d}\left(\boldsymbol{x}_{0}^{2}+\boldsymbol{y}_{0}^{2}\right)\right]\}$$\approx$1. The maximum intensity of the tweezer trap remains constant while the pattern of the tweezer array is randomly selected to generate a dataset.

\vspace{15pt}
\paragraph*{\bf  Generative neural network based on the U-net model}
The main idea of proposed scheme in this work is based on the U-Net structure that is a commonly used segmentation model providing contextual information~\cite{ronneberger2015u}. The U-Net architecture consists of a contracting down-sampling pathway that captures contextual information and a symmetric expanding up-sampling pathway that facilitates accurate localization (see Fig.~\ref{fig2})~\cite{li2022deep}. Notably, the incorporation of skip connections from the down-sampling to  up-sampling pathway constitutes a pivotal feature of the generative network, resulting in enhanced output image fidelity and detail retention. Here the down-sampling path is a residual neural network, consisting of down-sampling blocks and corresponding residual blocks. Each block is composed of a set of batch normalization, nonlinearity(ReLU), and a 4$\times$4 convolutional layer stacked one above the other. And the residual block effectively solves the degradation problem with skip connections~\cite{he2016deep}. The down-sampling is realized directly by convolutional layers with a stride of two and a padding of one. The up-sampling path consists of up-sampling blocks.

\begin{figure}
\centering
	\includegraphics[width=1\linewidth]{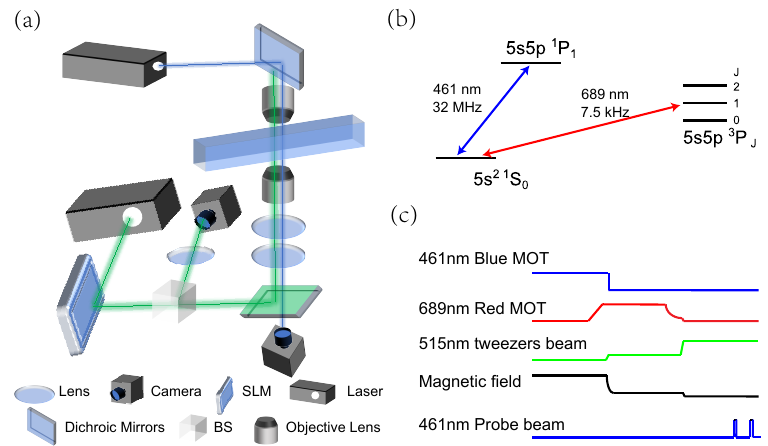}
	\caption{\textbf{Experiment scheme generating atomic arrays}  (a) A 515~nm tweezer beam generated by the SLM is projected to the glass cell through the bottom objective lens.  A 461-nm probe beam is directed from top to bottom and captured by a camera. (b) Energy-level diagram for bosonic strontium atom. Broad-line 461~nm and narrow-line 689~nm transitions are utilized to generate MOTs, respectively. (c) Sketch of the experimental sequence used to produce and load the tweezer arrays by projecting light from a SLM.}
	\label{fig3}
\end{figure}

\begin{figure}
\centering
	\includegraphics[width=1\linewidth]{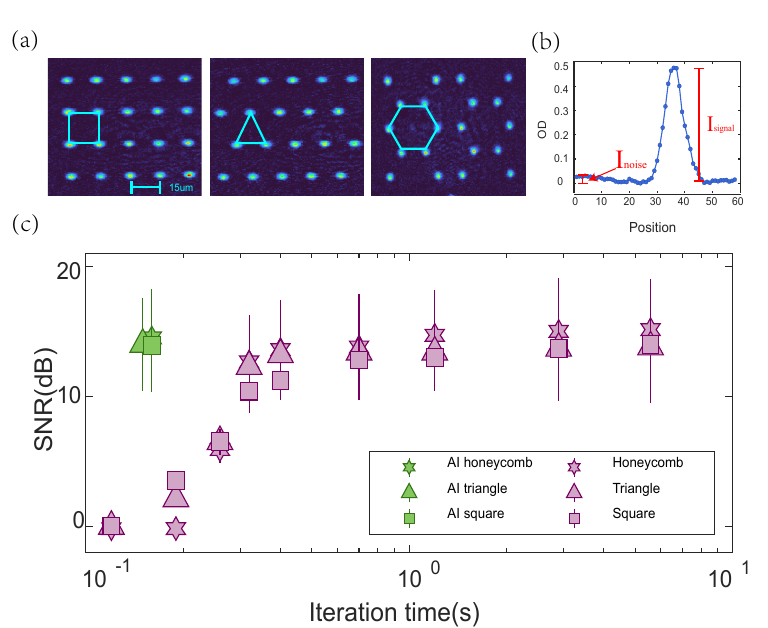}
	\caption{\textbf{Various atom arrays produced by the generative neural network}. (a)  Experimental absorption image of a $\sim$20 tweezer site, where each bright spot corresponds to an ensemble of $\sim$30 $^{88}$Sr atoms. Various lattices including square, triangular and honeycomb geometry are formed with the lattice constant of 15~$\mu$m. Absorption images are recorded with a CMOS camera.   (b) With the atomic signal at the center of the tweezer trap, we define the SNR as the ratio between $I_{signal}$ and $I_{noise}$. (c) Evaluation of algorithm runtime and tweezer array quality shows that the generative neural network outperforms the GS algorithm. The signal-to-noise ratio (SNR) of the tweezer array is achieved without requiring iteration. The error bar represents the  standard deviation.}
	\label{fig4}
\end{figure}

The trained generative network outputs the phase value which is to be similar to the desired phase pattern (i.e. POH) in the SLM plane. For training, we compare the output and the part of phase value in training dataset with Loss function $\mathcal{L}=\lvert \varphi_{out}-\varphi_{train} \rvert$. The generative network are implemented and trained using paddlepaddle 2.42~\cite{ma2019paddlepaddle} on an NVIDIA RTX 3080 GPU with Adam optimizer. The learning rate is 0.00001 and training runs for 6000 epochs.
The training of U-Net neural network takes 20 hours.

\vspace{10pt}
\paragraph*{\bf Loading strontium atoms into the tweezer array} 

A tweezer array generated by the computer-generated holography (CGH) based on the trained generative neural network is readily usable in a neutral atom experiment~\cite{wen.2024}. Here we produce a large sample of cold $^{88}$Sr atoms in a narrow-line magneto-optical trap (MOT) in a glass cell (see Fig.~\ref{fig3})~\cite{wen.2024,Yan.2024}. To obtain such a narrow-line MOT, a $^{88}$Sr atomic beam generated by an oven in a custom-designed  vacuum chamber is slowed down and captured in a broad-line 461~nm two-dimensional MOT. Subsequently, a 461~nm laser beam with low power, referred to as the "pushing beam" exerts a force on the atoms in the 2D-MOT, guiding them into a 461~nm 3D MOT in a separate glass cell as shown in Fig.~\ref{fig3}. Finally, we create a 689~nm narrow-line MOT in which a cold gas of 2$\times 10^6$ $^{88}$Sr atoms at the temperature of 2-5~$\mu$K are prepared at the center of the glass cell. The sample preparation process typically takes approximately 1 second~\cite{wen.2024}.
To create a tweezer array, programmable CGHs are generated by the use of phase SLMs (PLUTO-2.1-VIS-130) at the lens' back focal plane of the objective lens as described in Fig.~\ref{fig3}. A 515~nm laser beam of 0.5W  is sent to the SLM, resulting in about twenty tweezer traps with the trap depth of $k_B\times$1.2 mK where $k_B$ is the Boltzmann constant.

\vspace{15pt}
\paragraph*{\bf Experimental analysis} With a tweezer array being switched on, a few tens of Sr atoms are trapped in each micro trap, which allows us to examine the quality of the tweezer light pattern generated. To record the number of trapped atoms in each trap, a collimated 461nm laser beam is sent through the Sr atom array, and the optical density (OD) is recorded by the CMOS camera (Dhyana 400BSI V3)~\cite{wen.2024}. To avoid the unwanted reflections, such as stripes and Newtons rings, we employ a double-exposure scheme.  The first exposure is performed while the atoms are trapped in the optical tweezers within 10us while the second reference exposure is performed shortly after 0.5~s as a reference image. The OD is calculated according to $OD=ln\frac{I_{atom}-I_{ref}}{I_{atom}+I_{ref}}$ where $I_{atom}$ is the first image with atoms and $I_{ref}$ is the image without atom.

Fig.~\ref{fig4}(a) showcases a variety of arrays generated by the generative neural network. These arrays consist of square, triangular, and hexagonal lattices, with each tweezer trap containing tens of Sr atoms. In order to analyze these arrays, we extract the signal-to-noise ratio (SNR) from the absorption images shown in Fig.~\ref{fig4}(a). This is done by examining the maximum optical density at the center of each tweezer trap ($I_{signal}$) and comparing it to the background fluctuation ($I_{noise}$), as explained in Fig.~\ref{fig4}(b).

The main benefit of using the generative neural network for obtaining CGHs is that this approach overcomes the trade-off between computation time and the quality of the final atom arrays. The generative neural network outperforms the GS algorithm and the SNR of the tweezer array is achieved without requiring iteration. For example, the neural network can generate the POH in 0.16s with the SNR 14dB as shown in Fig.~\ref{fig4}(c). In contrast, atom arrays produced by the traditional GS algorithm face the problem of speckle noise due to the initial random phase and amplitude. As a result, several iterations are inevitably required to achieve a comparable SNR, which takes about 6 seconds.


In the existing architecture, there is a variation of approximately $\sim$10$\%$ in the trap depth of individual tweezers from site to site. However, this variation does not have a significant impact on the loading efficiency from the MOT. Nevertheless, we anticipate that precise control of the trap depth will be necessary when implementing laser cooling techniques~\cite{Tuchendler.2008}, such as Raman sideband or Sisyphus cooling. To produce a more uniform atom array and further reduce the computational complexity, many extensions to the current method are worth investigating. One is training the generative network with weighted GS algorithm~\cite{di2007computer}. For every atom array point, weighted coefficient feedback from every iteration will enhance uniformity. The current network architecture is based on the universal U-Net. It is conceivable that other advanced network architectures, such as generative adversarial network and graph neural network can also be applicable.

\vspace{15pt}
\paragraph*{\bf Conclusion} In summary, we have introduced a method for generating tweezer arrays to trap cold atoms using state-of-the-art generative neural networks. Our findings demonstrate the remarkable capability of the U-net based deep learning approach in successfully producing arrays of Sr atoms. Importantly, this method eliminates the need for time-consuming iterative procedures such as the GS algorithm. Furthermore, the Sr atom array discussed in this work is not limited to its current form but can be easily applied to fermionic isotopes. This would open up exciting possibilities for exploring the SU($N$) physics in a tweezer trap~\cite{He.2019}.

\vspace{15pt}
\paragraph*{\bf Data Availability}
The data that supports the findings of this study are available within the article. 

\vspace{7pt}
\paragraph*{\bf Acknowledgement} GBJ acknowledges support from the RGC through 16306119, 16302420, 16302821, 16306321, 16306922, 16302123, C6009-20G, N-HKUST636-22, and RFS2122-6S04. RZJ acknowledges support from the CPSF through 2022M720890. CH acknowledges support from the RGC for RGC Postdoctoral fellowship.

\bibliography{allref.bib}

\end{document}